\documentclass[journal,twocolumn]{IEEEtran}

\usepackage{amsfonts}
\usepackage{times}
\usepackage{graphicx}
\usepackage{latexsym}
\usepackage{dsfont}
\usepackage{amssymb}
\usepackage{amsmath}
\usepackage{cite}
\usepackage{verbatim}
\usepackage{subfigure}

\newcommand{\figref}[1]{{Fig.}~\ref{#1}}

\def\bb0{{\mathbb{0}}}

\def\bb{{\mathbf{b}}}

\def\bd{{\mathbf{d}}}

\def\bff{{\mathbf{f}}}

\def\b0{{\mathbf{0}}}

\def\bS{{\mathbf{S}}}

\def\bX{{\mathbf{X}}}

\def\bbE{{\mathbb{E}}}

\def\sf0{{\mathsf{0}}}

\usepackage{epstopdf}
\usepackage{enumerate}
\usepackage{algorithmicx}
\usepackage{algorithm}
\usepackage{amsmath}
\usepackage[noend]{algpseudocode}
\usepackage{float}
\usepackage{hyperref}
\usepackage{color}
\usepackage{makeidx}
\usepackage{bbm}
\usepackage{graphicx}
\usepackage{lipsum}
\usepackage{subfigure}
\usepackage{tablefootnote}
\usepackage{multirow}
\usepackage{multicol}
\usepackage{balance}
\usepackage{booktabs}
\usepackage{soul}
\usepackage[normalem]{ulem}
\usepackage{xcolor}

\newcommand{\sref}[1]{{Section}~\ref{#1}}

\newcommand{\comm}[1]{}
\begin{document}

\title{Computer Vision Aided Blockage Prediction in  Real-World Millimeter Wave Deployments}
\author{Gouranga Charan and  Ahmed Alkhateeb\\ \textit{School of Electrical, Computer and Energy Engineering - Arizona State University} \\ \textit{Emails: \{gcharan, alkhateeb\}@asu.edu} }

\maketitle

\begin{abstract}

This paper provides the first real-world evaluation of using visual (RGB camera) data and machine learning for proactively predicting millimeter wave (mmWave) dynamic link blockages before they happen. Proactively predicting line-of-sight (LOS) link blockages enables mmWave/sub-THz networks to make proactive network management decisions, such as proactive beam switching and hand-off) before a link failure happens. This can significantly  enhance the network reliability and latency  while efficiently utilizing the wireless resources. To evaluate this gain in reality, this paper (i) develops a computer vision based solution that processes the visual data captured by a camera installed at the infrastructure node and (ii) studies the feasibility of the proposed solution based on the large-scale real-world dataset, DeepSense 6G, that comprises multi-modal sensing and communication data. Based on the adopted  real-world dataset, the developed solution achieves $\approx 90\%$ accuracy in predicting blockages happening within the future $0.1$s and $\approx 80\%$  for blockages happening within $1$s, which highlights a promising solution for  mmWave/sub-THz communication networks.

\end{abstract}


\section{Introduction} \label{sec:Intro}

Millimeter wave (mmWave) and sub-terahertz communication systems  rely on line-of-sight (LOS) links to achieve sufficient receive signal power. Blocking these LOS links by the moving objects in the environment may disconnect the communication session or cause sudden and significant degaradtion in the link quality. This is due to the high penetration loss of the mmWave/sub-terahertz signals and the much less receive power of the NLOS links compared to the LOS ones \cite{Rappaport2019,Andrews2017}. All that highly challenges the reliability and latency of the mmWave/sub-terahertz communication networks. Initial approaches for overcoming these blockage challenges relied mainly on multi-connectivity \cite{Polese2017,Petrov}. These solutions, however, generally keep the user connected to multiple infastructure nodes which underutilizes the wireless network resources. This motivated the research for more efficient blockage avoidance approaches. 

Leveraging machine learning (ML) to address the blockage challenges has gained increasing interest in the last few years \cite{Alkhateeb2018,wu2021blockage, Alrabeiah2020}. In \cite{Alkhateeb2018}, the authors proposed to leverage recurrent neural networks to process the sequence of beams serving a mobile user and to predict whether or not a future blockage will happen. Relying only on beam sequences, however, limits the applications to stationary blockage prediction. Predicting dynamic blockages require more information about these moving blockages in the environment. In \cite{wu2021blockage,Alrabeiah2020},  in-band mmWave and sub-6GHz based wireless scattering signatures were used to indentify/predict the incoming mmWave link blockges. These solutions, however, are mainly capable of predicting immediate blockages and are hard to scale to complex/crowded scenarios. To enable predicting blockages early enough before they block the links, solutions based on radar and LiDAR sensory data were proposed for the first time in \cite{demirhan2021radar_ICC, wu2021lidar_WCNC}. Despite their promising results, easy sensing modality has its advantages and drawbacks. For examlpe, radar data are mainly suitable for uncrowded scenarios and LiDAR sensors are expensive and have relatively short range. 

\begin{figure}[!t]
	\centering
	\includegraphics[width=1.0\linewidth]{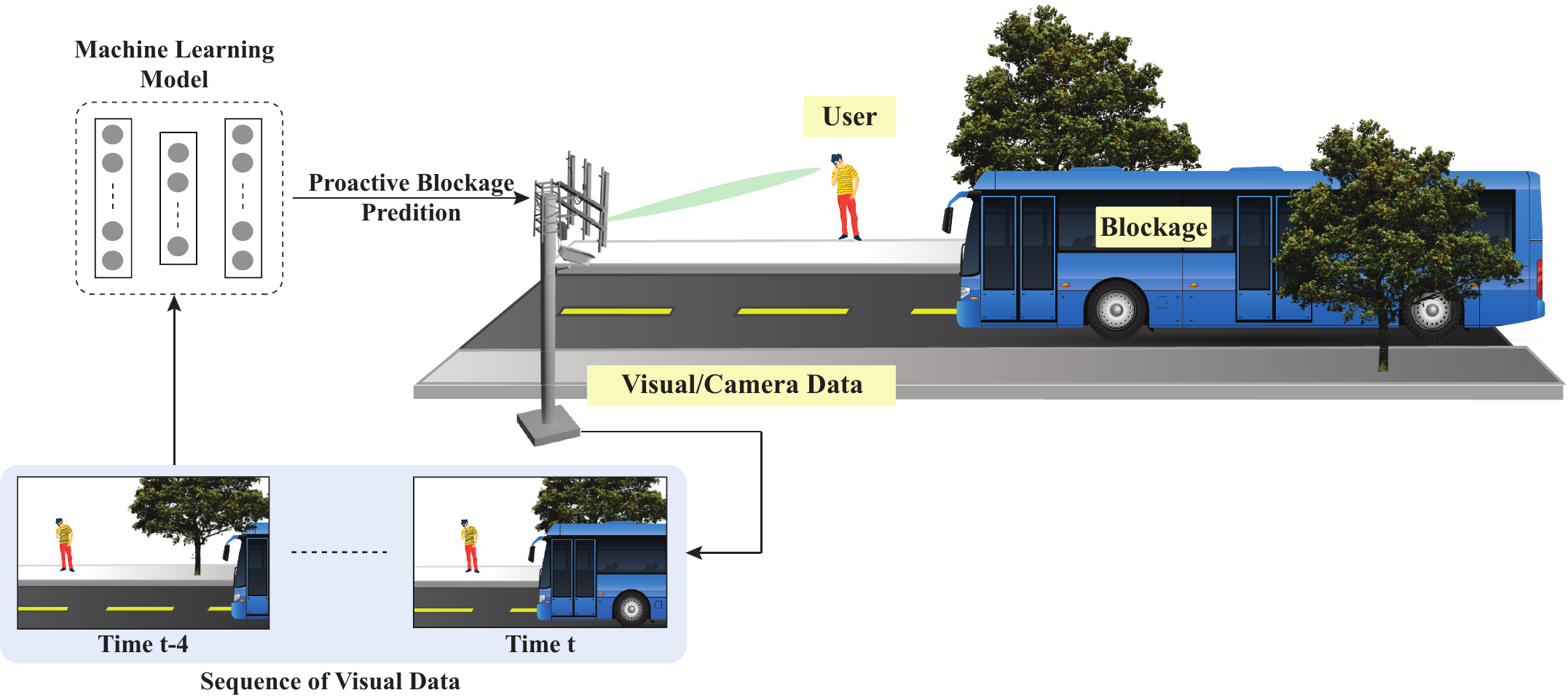}
	\caption{This figure illustrates the overall system model where a mmWave/THz basestation utilizes the captured RGB images to predict the future link blockage status. }
	\label{fig:key_idea}
	\vspace{-3mm}
\end{figure}

In \cite{Charan2021b}, we proposed to leverage visual data (captured by cameras) to predict future dynamic blockages. Ths solutions and analysis in \cite{Charan2021b}, though, were based only on synthetic datasets, and an important question that arises is whether the promising results in \cite{Charan2021b} can be achieved in reality? In this paper, we attempt to answer this question. In particular, the main contributions of the paper can be summarized as follows:
\begin{itemize}
	\item Formulating the vision-aided blockage prediction problem in mmWave/THz wireless networks considering practical visual and communication models. 
	\item Developing a machine learning approach that is capable of (i) pre-processing the real-world visual data to enhance the blockage prediction performance, (ii) extracting the relevant features about the scatterers/environment, and (iii)  efficiently predicting future dynamic link blockages.  
	\item Providing the first real-world evaluation of vision-aided blockage prediction based on our large-scale dataset, DeepSense 6G \cite{DeepSense}, that consists of co-existing multi-modal sensing and  wireless communication data. 
\end{itemize}
Based on the adopted real-world dataset, the developed solution achieves $\approx 90\%$ accuracy in predicting blockages happening within a future prediction interval of $0.1$s and $\approx 80\%$  for a prediction interval of $1$s. This highlights the potential of leveraging machine learning and visual data in addressing the critical LOS link blockage challenges. In particular, the capability to efficiently predict future blockages enable the network to make proactive beam/basestation switching decisions which enhances the overall network reliability/latency performance.

\section{System Model }\label{sec:sys_ch_mod}

This work considers a communication scenario where a mmWave basestation is serving a stationary user located in a busy environment with multiple moving objects, such as vehicles, pedestrians, etc., as shown in \figref{fig:key_idea}. The mmWave basestation is equipped with an RGB camera to monitor and gather sensing data about the surrounding environment. This information could potentially be leveraged to proactively predict future link blockages caused by the moving objects. 

The adopted  system model consists of a mmWave basestation equipped with an $N$-element antenna array and a standard-resolution RGB camera. The basestation is serving a stationary user that is, for simplicity, considered to have a single antenna. The basestation uses a pre-defined beam codebook $\boldsymbol{\mathcal F}=\{\mathbf f_m\}_{m=1}^{M}$ to serve the user, where $\mathbf{f}_m \in \mathbb C^{N\times 1}$ and $M$ is the total number of beamforming vectors in the codebook. As will be described in Section~\ref{sec:testbed}, the beamforming codebook adopted by the hardware prototype has $64$ beamforming vectors (i.e., $M=64$) with the azimuth angles uniformly quantized between $\left[-\frac{\pi}{4}, \frac{\pi}{4} \right]$. The communication system further adopts OFDM transmission with $K$ subcarriers and cyclic prefix of length $D$. At any time instant $t$, if the basestation uses the beamforming vector $\mathbf f_m \in \boldsymbol{\mathcal F}$ to serve the user, then the downlink received signal at the user at the $k$th subcarrier can be expressed as
\begin{equation}
    y_{k}[t] = \mathbf h_{k}^T[t] \bff_m x[t] + n_{k}[t],
    \label{eq:received_signal}
\end{equation}
where $\mathbf h_{k}[t] \in \mathbb C^{N \times 1}$ is the channel between the basestation and the user at the $k$th subcarrier, $x[t] $ is a transmitted data symbol, $\bbE{\left|x[t]\right|^2}=P$, with the average transmit power $P$, and  $n_k$ is a receive noise sample,  $n_k \sim \mathcal N_\mathbb C(0,\sigma_n^2)$.

\textbf{LOS Blockage:} The channel model $\mathbf h_{k}$, defined in \eqref{eq:received_signal}, is generic and can be expressed as follows at time instant $t$
\begin{equation}
	\mathbf h_{k}[t] = \left( 1 - b[t]\right) \mathbf h_{k}^{\text{LOS}}[t] + \mathbf h_{k}^{\text{NLOS}}[t],
\end{equation} 
where $\mathbf h_{k}^{\text{LOS}}$ and $\mathbf h_{k}^{\text{NLOS}}$ are the LOS and NLOS channel components. The binary variable $ b[t] \in [0,1] $ represents the link status at time instant $t$, with $b[t] = 1$ indicating that the LOS path is blocked and $b[t] = 0$ otherwise. 

It is important to note here that for mmWave and sub-THz communication systems, the LOS channel gain is much greater than the NLOS channel gain \cite{ Rappaport2019,Andrews2017}. Therefore, LOS link blockages challenge the reliability of these networks. Next, we provide a formal definition of the  proactive vision-aided blockage prediction problem which is the focus of this work.

\section{Vision-Aided Blockage Prediction: \\ Key Idea and Problem Formulation} \label{sec:key_idea_prob_form}

%In this section, we build upon the system model presented in Section~\ref{sec:sys_ch_mod} to formally define the vision-aided blockage prediction problem. 

One of the major challenges in the high-frequency wireless communication networks is the LOS link blockages; the mmWave/THz communication systems suffer from link disconnection and significant dips in the received SNR when an object/blockage intersects the LOS path between the basestation and the user. Re-establishing a LOS connection is usually done in a reactive way, which incurs critical latency and impacts the reliability of such systems. The presence of dynamic moving objects in the environment further increases these reliability/latency challenges. These challenges could potentially be addressed if these blockages can be proactively predicted \cite{Charan2021b, wu2021c, demirhan2021radar}. In order to develop an efficient solution that can \textit{proactively} predict the occurrence of such future blockages, it is essential to equip the wireless network with a sense of its surroundings. This work attempts to do so by utilizing machine learning and visual data captured by cameras placed at the basestation to proactively predict future blockages before they happen. In this section, we will first present the key idea in Section~\ref{sec:key_idea} and then formulate  the vision-aided blockage prediction problem in Section~\ref{sec:prob_form}.

%####################################################################################################

\begin{figure*}[!t]
	\centering
	\includegraphics[width=0.95\linewidth]{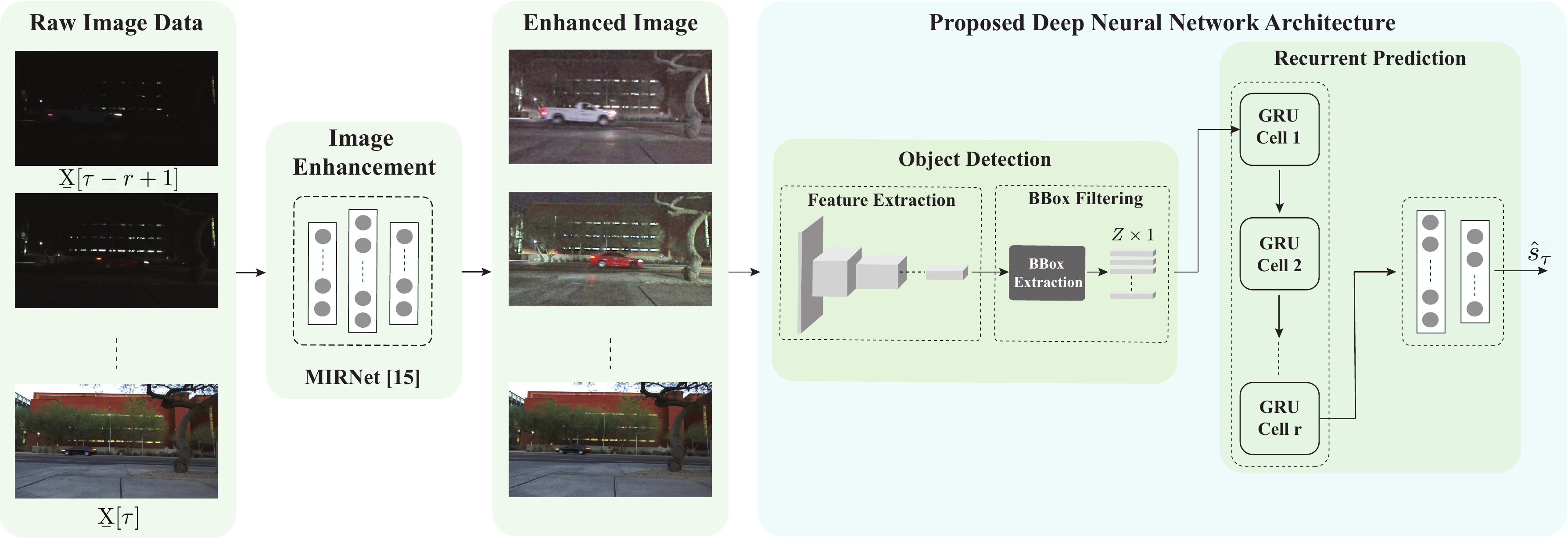}
	\caption{This figure presents the proposed architecture for vision-aided future blockage prediction. The raw visual data is first processed to enhance the images captured during night or low-light scenario. The processed sequence of images are then passed through the proposed deep neural network architecture consisting of a object detection model and recurrent neural network to predict the future blockage.  }
	\label{fig:ML_main_fig}
	%\vspace{-4mm}
\end{figure*}
%####################################################################################################

\subsection{The Key Idea}\label{sec:key_idea}

In a wireless network, the link blockages are often caused by moving objects, such as cars, trucks, buses, and humans, present in the wireless environment. Given the dynamic nature of multiple moving objects in a real-wireless scenario, the task of future blockage prediction becomes extremely challenging. While the detection of the different objects in the environment, such as cars and humans, can be well achieved using a single image, the success of the dynamic blockage prediction task also relies on characterizing the mobility patterns and geometric features of these objects. For example, for a vehicle-to-infrastructure use case, the visual data at the infrastructure needs to characterize the speed/direction of travel and the size of the various objects in the environment. In order to capture these additional indicators, which are normally obtained by analyzing a \textit{sequence} of images, \textbf{our proposed solution observes a sequence of $r$ image samples instead of making the predictions based on just one sample and attempts to predict future LOS link blockages before they happen}. Building upon the key idea presented here, in the next sub-section, we provide the formal definitions for the proactive vision-aided blockage prediction problem.

\subsection{Problem Formulation} \label{sec:prob_form}
The main objective of this work is to observe a sequence of camera image samples captured at the basestation and utilize the sensing data to predict whether or not the stationary user will be blocked within a window of future instances. Let $\bX[t] \in \mathbb{R}^{W \times H \times C} $ denote a single RGB image of the environment captured at the basestation at time instant t, where $W$, $H$, and $C$ are  the width, height, and the number of color channels for the image. At any time instant $\tau\in \mathbb Z$, the basestation uses a sequence of RGB images, $\bS[\tau]$, defined as
\begin{equation}
	{\bS}[\tau] = \left\{ \bX[t] \right\}_{t = \tau-r+1}^{\tau}, 
\end{equation}
where $r \in \mathbb Z$ is the length of the input sequence or the observation window to predict future link blockages.  In particular, at any given time instant $\tau$, the goal in this work is for the basestation to observe $\bS[\tau]$ and predict whether or not the stationary user is going to be get blocked within a window of $r^\prime$ future instances. It is important to note here that we do not focus on the exact future instance but consider the entire future window sequence for denoting the future blockage status. Given ${\bS}[\tau]$ and the future window $r^\prime$, the future blockage status at time instant $\tau$ can then be expressed as 
\begin{equation}\label{eq:ls}
	s [\tau] = \left\{ \begin{array}{ll}
		1, & b[t] = 1,\  t\in\{\tau+1,\dots,\tau+r^{\prime}\}  \\
		0, & \text{otherwise} \\
	\end{array}
	\right.
\end{equation}
where $0$ indicates that the user remains LOS within the next $r^\prime$ future instances and $1$ points towards the occurrence of blockage within the same window. 

In order to predict the future blockage status, we define a function $f_{\Theta}$ that {maps} the observed sequence of images, $\bS[\tau]$ to a prediction (estimate) of the future blockage status, $\hat{s}_{\tau}$. The function $f_{\Theta}$ can be formally expressed as 
\begin{equation}
	f_{\Theta}: \bS[\tau] \rightarrow \hat{s}[{\tau}].
\end{equation}
In this work, we adopt a machine learning model to learn this prediction function $f_{\Theta}$, that takes in the observed image sequence and predicts the future blockage status, $\hat s[{\tau}] \in \{0, 1\}$. Here, $\Theta$ represents the parameters of the machine learning model and is learned from a dataset of labeled sequences. For this, a dataset of independent sample pairs $\mathcal D = \left\{ \left(\bS_{v}, s_{v}\right) \right\}_{v = 1}^{V}$ is collected, where $s_{v}$ is the ground-truth future blockage label for the observed sequence $\bS_{v}$, and $V$ is the total number of \textit{sequence-label} pairs in the dataset. The labeled dataset $\mathcal D$ is then used to optimize the prediction function $f_{\Theta}$ such that it maintains high fidelity for any samples drawn from this dataset. The optimization problem can be  written as
\begin{equation}\label{eq:joint_prob}
f^\star_{\Theta^\star}	 = \underset{f_{\Theta}(.)}{\text{argmax}} \quad \prod_{v=1}^{V} \mathbb P(\hat s_v = s_v|\bS_v),	
\end{equation}
where the joint probability in \eqref{eq:joint_prob} is factored out to convey the identical and independent (i.i.d.) nature of the samples in dataset $\mathcal D$. In the next section, we present the proposed deep learning-based solution for the vision-aided future blockage prediction task.

\section{Vision-Aided Blockage Prediction: \\ A Deep Learning Solution} \label{sec:prop_sol}

Guided by the principles mentioned in Section~\ref{sec:key_idea}, the blockage prediction task is divided into two sub-tasks: (i) object detection and (ii) recurrent prediction. The first sub-task deals with detecting the relevant objects of interest in the FoV of the basestation. Given the recent advancements in the field of computer vision and deep learning, this task can be performed by utilizing convolutional neural network (CNN)-based object detectors such as the You Only Look Once (YOLO) model \cite{Yolov3}. The first stage consisting of the object detection models extract relevant features from the sequence of images and provides this as an input to the next stage of the machine learning pipeline. The objective of the second stage needs is to predict the future blockages based on these extracted features. Recurrent Neural Networks (RNNs) are state-of-the-art machine learning models specifically designed to deal with such sequential learning problem. Therefore, we adopt a recurrent neural network in the second stage to learn the underlying key indicators from the extracted features and predict the future link blockages. In Fig.~\ref{fig:ML_main_fig}, we illustrate the proposed deep learning-based blockage prediction solution. In this section, we first present the details of the image enhancement pre-processing stage adopted to deal with the low-light/dark images. Then, we take a deeper dive into the developed two-stage vision-aided blockage prediction solution.

\subsection{Data Processing (Image Enhancement)}
Compared to LiDAR, radars and other sensing modalities, RGB cameras provide a low-cost, high-resolution, and low-footprint alternative, making it one of the preferred choices for wireless sensing applications. However, there is major bottleneck associated with the visual images captured using an RGB camera. Under low-light conditions, the visual data turns out to be noisy and dark, making it unsuitable for further computer vision tasks. Fig.~\ref{fig:ML_main_fig} shows an image captured under such low light conditions. The white truck in the first image and the red sedan in the second are hardly visible highlighting the challenges associated with such images. In order to develop a robust and reliable solution that can work in most of the natural lighting conditions, it is essential to perform some sort of image enhancement to extract the hidden details and make the low-light images more usable. For the post-processing stage, we adopt the state-of-the-art MIRNet \cite{MIRNet} model developed for low-light image enhancement. It is a fully-convolutional architecture that learns an enriched set of features by combining contextual information from multiple scales, while simultaneously preserving the high-resolution spatial details. As shown in Fig.~\ref{fig:ML_main_fig}, that the objects in the low-light images are clearly visible after the image enhancement post-processing step, which is important for the performance of the proposed blockage prediction solution. 
%####################################################################################################
\begin{figure}[!t]
	\centering
	\includegraphics[width=1.0\linewidth]{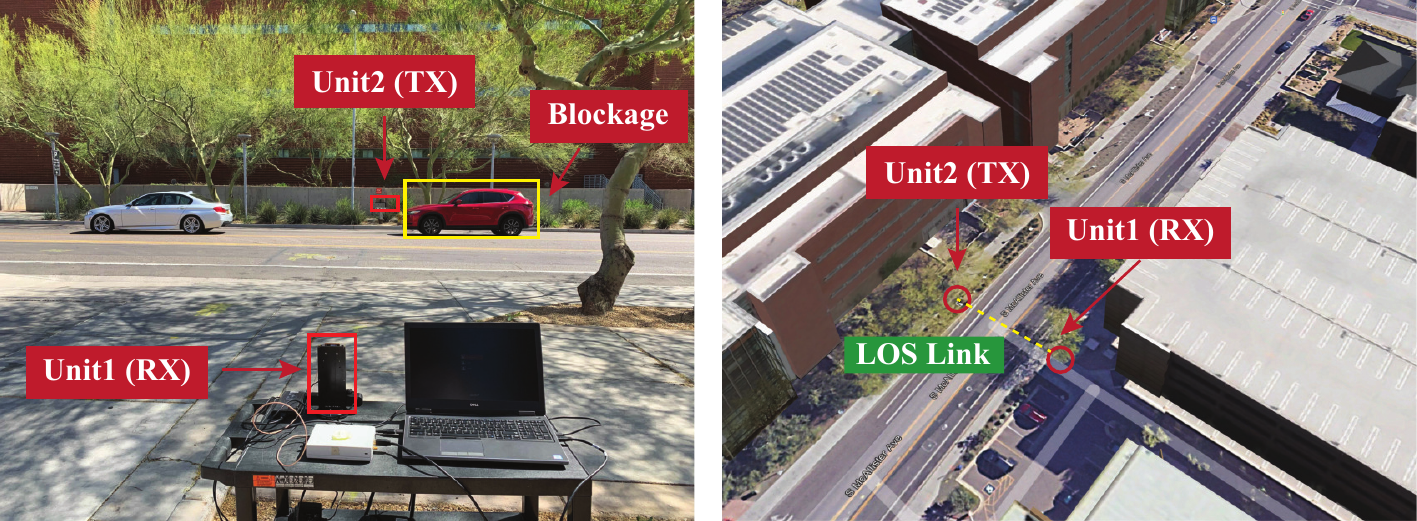}
	\caption{This figure shows the data collection setup used for DeepSense 6G Scenarios 17-22. The figure on the left depicts the street view from Unit 1 perspective. It also highlights the location of the transmitter (Unit 2 TX) and the receiver (Unit 1 RX) during the data collection process. The figure on the right shows the exact location of Unit 1 and Unit 2 highlighted on the Google Earth 3D view. }
	\label{fig:deepsense}
	\vspace{-4mm}
\end{figure}
%####################################################################################################

\subsection{A Two-Stage Deep Learning Model} Here, we present the details of the proposed blockage prediction architecture, which consists of two key functions, namely object detection and recurrent prediction. 

\textbf{Object Detection: }The first stage of the proposed solution is the object detection deep learning model. There are two primary goals of this stage: (i) Perform accurate and quick detection of the objects of interest in the FoV of the basestation and (ii) extract the coordinates of the bounding boxes placed around the relevant objects. For this, in our proposed solution, we adopt the state-of-the-art YOLO object detection model and more specifically the further improved YOLOv3 architecture \cite{Yolov3}. The YOLOv3 detector is a fast and reliable end-to-end object detection system, designed for real-time processing of visual data (images and videos). In this work, instead of training the YOLOv3 object detection model from scratch, we utilize the COCO pre-trained model as it is already capable of detecting most of the relevant objects present in a wireless environment. 

The pre-trained YOLOv3 architecture is particularly selected for the bounding box detection task in this work. For each image sample, the pre-trained YOLOv3 is used to detect the relevant objects and extract the bounding box coordinates of the detected objects. In particular, for each detected object in the image, we extract a $4$-dimensional vector consisting of the bottom-left coordinates $[x_1, y_1]$ and the top-right coordinates $[x_2, y_2]$. These coordinates are normalized to be between $[0,1]$. In order to account for multiple detected objects in the FoV of the basestation, the extracted bounding boxes are concatenated to form one dimensional vector $\bd\in \mathbb R^{Y \times 1}$, where $Y$ is the number of objects detected by the YOLOv3 model. It is important to highlight here that the number of detected objects might not be the same in each data sample, which  results in a variable length vector $\bd$. This will lead to inconsistency in the size of the extracted features and create unnecessary complications for the next stage of the proposed solution pipeline, i.e., the recurrent predictions. In order to avoid this inconsistency, the extracted bounding box vector $\bd$ is further padded with $Z-Y$ zeros to obtain a fixed size vector $\tilde \bd \in \mathbb R^{Z \times 1}$. The fixed size bounding box feature vector $\tilde \bd$ is then provided as an input to the recurrent network to predict the future link blockage status.

\textbf{Recurrent Prediction:} The final stage of the proposed solution utilizes recurrent neural networks to make the final prediction. In this work, we consider a two-stage Gated Recurrent Unit (GRU), followed by a fully-connected layer acting as a classifier. More specifically, the model receives  a sequence of $r$ extracted bounding box feature vectors, $\{\tilde \bd[\tau-r+1, \ldots, \tilde \bd[\tau]]\}$, as input and predicts the future link blockage status over a window of $r^{\prime}$ time instance, $\hat{s}[\tau]$.

%####################################################################################################
\begin{figure*}[!t]
	\centering
	\includegraphics[width=0.75\linewidth]{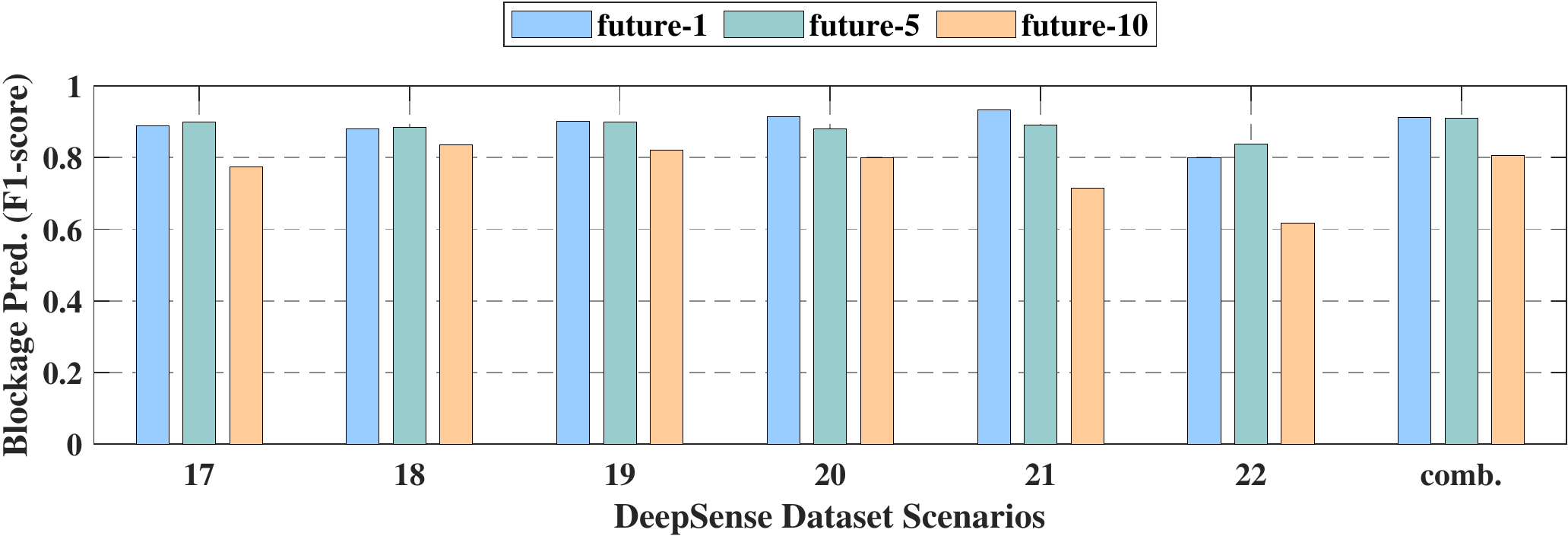}
	\caption{This figure plots the future-1, future-5, and future-10 future blockage prediction scores (f1-score) for scenarios 17-22 and the combined scenarios. The combined scenario achieves comparable or better prediction accuracy highlighting the gain of having sufficient diversity in the dataset.}
	\label{fig:acc_bar_plot}
	\vspace{-4mm}
\end{figure*}
%####################################################################################################

%###########################################################################################

\begin{table}[!t]
	\caption{Number of Data Sequences in the Development Dataset}
	\centering
	\setlength{\tabcolsep}{5pt}
	\renewcommand{\arraystretch}{1.2}
	\begin{tabular}{|c|ccc|c|}
		\hline
		\multirow{2}{*}{\textbf{\begin{tabular}[c]{@{}c@{}}DeepSense 6G \\ Scenarios\end{tabular}}} & \multicolumn{3}{c|}{\textbf{Number of Sequences}}                                        & \multirow{2}{*}{\textbf{\begin{tabular}[c]{@{}c@{}}Time of\\ the day\end{tabular}}} \\ \cline{2-4}
		& \multicolumn{1}{c|}{\textbf{Train}} & \multicolumn{1}{c|}{\textbf{Val.}} & \textbf{Test} &                                                                                     \\ \hline \hline
		17                                                                                          & \multicolumn{1}{c|}{4176}           & \multicolumn{1}{c|}{1190}          & 638           & Day + Night                                                                         \\ \hline
		18                                                                                          & \multicolumn{1}{c|}{4524}           & \multicolumn{1}{c|}{1276}          & 667           & Day + Night                                                                         \\ \hline
		19                                                                                          & \multicolumn{1}{c|}{11455}          & \multicolumn{1}{c|}{3277}          & 1653          & Day + Night                                                                         \\ \hline
		20                                                                                          & \multicolumn{1}{c|}{5307}           & \multicolumn{1}{c|}{1508}          & 783           & Day                                                                                 \\ \hline
		21                                                                                          & \multicolumn{1}{c|}{1131}           & \multicolumn{1}{c|}{319}           & 203           & Day                                                                                 \\ \hline
		22                                                                                          & \multicolumn{1}{c|}{377}            & \multicolumn{1}{c|}{110}           & 55            & Day                                                                                 \\ \hline
	\end{tabular}
	\label{tab_num_samples}
	\vspace{-4mm}
\end{table}

%##############################################################################################

\section{Testbed Description and Development Dataset}\label{sec:datset}

In order to evaluate the performance of the proposed vision-aided blockage prediction solution, we adopt multiple scenarios from the DeepSense 6G \cite{DeepSense} dataset. DeepSense 6G is a real-world multi-modal dataset enabling sensing-aided wireless communication applications. It contains co-existing multi-modal data such as vision, mmWave wireless communication, GPS data, LiDAR, and Radar, collected in realistic wireless environments. In this section, we present a brief overview of the scenarios adopted from the DeepSense 6G dataset followed by the analysis of the final development dataset utilized in the blockage prediction task. 

%###########################################################################################

\begin{table}[!t]
	\caption{Design and Training Hyper-parameters}
	\label{table}
	\centering
	\setlength{\tabcolsep}{5pt}
	\renewcommand{\arraystretch}{1.2}
	\begin{tabular}{|l|c|c|}
		\hline
		\multirow{4}{*}{Design}   
		& Number of GRUs Per Layer ($r$)   & $8$             \\  \cline{2-3}
		& Embedding Dimension ($Z$)        & $30$                        \\ \cline{2-3} 
		& Hidden State Dimension         & $128$                         \\ \cline{2-3} 
		& Number of classes  & $2$                        \\ \hline\hline 
		%                          & Dropout Percentage             & 0.2                        \\ \hline
		\multirow{5}{*}{Training} & Optimizer                      & ADAM                       \\ \cline{2-3} 
		& Learning Rate                  & $1 \times 10 ^{-3}$ \\ \cline{2-3} 
		& Batch Size                     & $128$                       \\ \cline{2-3} 
		& Number of Training Epochs               & $100$                         \\ \hline
	\end{tabular}
	\label{tab:design_param}
	\vspace{-4mm}
\end{table}

%##############################################################################################

\textbf{DeepSense 6G: [Scenarios 17 - 22]}  \label{sec:testbed}We adopt Scenarios 17-22 of the DeepSense 6G dataset for evaluating the efficacy of our proposed solution. The hardware testbed and the exact location used for collecting these data is shown in Fig.~\ref{fig:deepsense}. The DeepSense testbed $3$ is utilized for this data collection and is placed on the opposite sides of a $2$-way street with a passing-lane in-between. The primary components of the adopted testbed are: (i) A stationary 60 GHz omni-directional mmWave transmitter (unit2), (ii) a directional mmWave receiver (unit1), (iii) an RGB camera. The receiver employs a 16-element ($N = 16$) 60 GHz phased array and it receives the transmitted signal using an over-sampled beam codebook of $64$ pre-defined beams ($M = 64$). Unit 1 is also equipped with a camera and it captures RGB images for the wireless environment in the FoV of the receiver. The testbed captures data at 12 samples/sec for scenarios 17-19 and 6.5 samples/sec for scenarios 20-22. Each data sample consists of an RGB image of the environment and a 64-element mmWave receive power vector. For more information regarding the data collected testbed and setup, please refer to \cite{DeepSense, wu2021blockage}.

\textbf{DeepSense 6G: [Development Dataset]} \label{sec:dev_data} The adopted DeepSense scenarios include diverse data collected during different times of the day (day and night). Each row in the dataset scenarios consists of a tuple of an RGB image, $\bX[\tau]$, and the corresponding receive power vector and the ground-truth link blockage status, $s[{\tau}]$ (manually labelled).  To form the development dataset of the blockage prediction task described in \sref{sec:prob_form}, the offered DeepSense data is further processed using a sliding window to generate a time-series dataset consisting of $8$ input image samples ($r = 8$) and the corresponding future blockage status in a future window  of $r^\prime$ samples (we generate 10 such time-series datasets at $r^\prime = 1, 2, ..., 10$.  In Table~\ref{tab_num_samples}, we present the number of such sequences in each scenario. The numbers shown in the table corresponds to an input sequence length of $8$ and the future prediction window length of $10$. In order to perform an in-depth study, the development dataset per scenario was further processed to generate datasets for different future prediction window size. For example, \textit{future-1} dataset, consists of data sequences where the input sequence length is still $8$, but the future prediction window length is $1$. During these process of generating datasets with different future prediction window, we ensure that the dataset is balanced \{Number of LOS data sequences $\approx$ Number of NLOS sequences.\}. Each of the development dataset is further divided into training, validation and test sets following a split of $70-20-10\%$.  

%####################################################################################################

\begin{figure}[!t]
	\centering
	\includegraphics[width=0.85\linewidth]{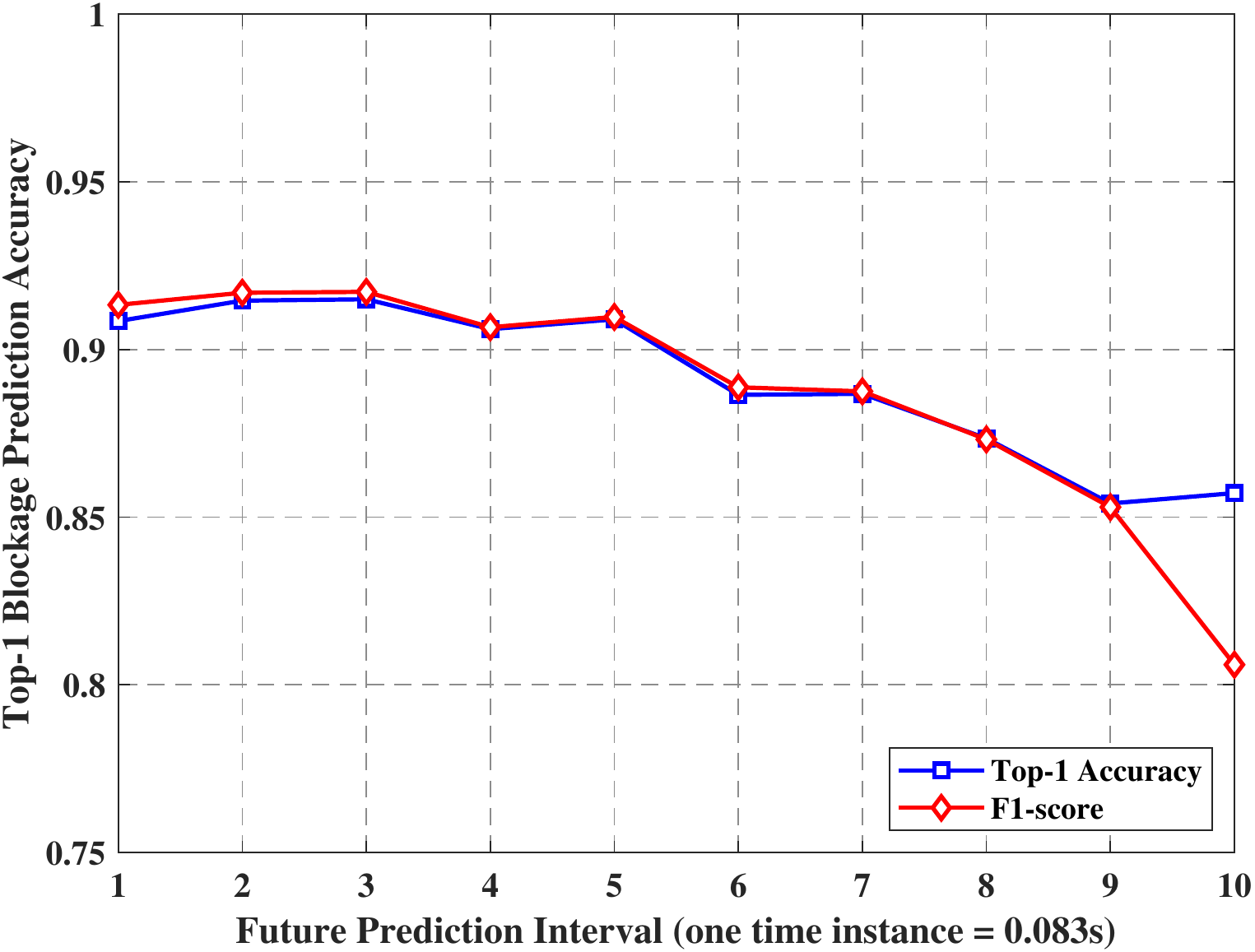}
	\caption{Top-1 blockage prediction accuracy and F1-score for different future prediction intervals observed in the combined case. It is observed that both the top-1 accuracy and f1-score decreases as we predict further into the future.}
	\label{fig:combined_scene_score}
	%\vspace{-4mm}
\end{figure}

\section{Performance Evaluation} \label{sec:perf_eval}
In this section, we first discuss the neural network training parameters and the adopted evaluation metrics. Next, we present the numerical evaluation of the proposed solution.

\textbf{Experimental Setup:} As described in Section~\ref{sec:prop_sol}, this work adopts a pre-trained object detection model (YOLOv3) to first extract the coordinates of the bounding boxes from a sequence of image data. The extracted bounding boxes are then provided as an input to the recurrent neural network proposed earlier in Section~\ref{sec:prop_sol}. The GRU model is trained using the labeled development dataset discussed in Section~\ref{sec:dev_data} using a cross-entropy loss function. All the simulations were performed on a single NVIDIA Quadro 6000 GPU using the PyTorch deep learning framework. The detailed design and training hyper-parameters are presented in Table~\ref{tab:design_param}. We utilize the top-1 accuracy metric as the primary method of evaluating the proposed solution. The top-1 accuracy is defined as follows:
\begin{equation}
	Acc_{top-1} = \frac{1}{U} \sum_{u=1}^{U} \mathbbm{1} \{\hat s_u[\tau] = s_u[\tau] \},
\end{equation}
where $\hat s_u[\tau]$ and $s_u[\tau]$ are the predicted and ground-truth link blockage status, respectively. $U$ is the total number of samples present in the validation/test set. $\mathbbm{1}\{.\}$ is the indicator function.  In order to study the robustness of the proposed solution, we also utilize the F1-score metric.

\textbf{Can visual data predict LOS blockages?} For each of the scenarios 17-22, we evaluate the proposed solution for various future prediction window lengths. In Fig.~\ref{fig:acc_bar_plot}, we show the F1-score of the future-1, future-5 and future-10 blockage prediction. Note that `future-5' here represents the development dataset that considers a  future blockage prediction window of length $5$ time instances (i.e., predicting a blockage that will happen in the future 450ms in scenarios 17-19 and 750ms in scenarios 20-22). In Fig.~\ref{fig:acc_bar_plot}, we observe that for all the scenarios, {the proposed two-staged solution achieves an accuracy of $\approx 0.88 - 0.90$ future-1 and future-5 blockage prediction F1-score}, highlighting the high efficiency of the proposed vision-aided blockage prediction approach. It is observed that there is a slight degradation in the model's performance for scenario 22, which could be attributed to the lower number of samples in the development dataset of this scenario as shown Table~\ref{tab_num_samples}; the lower number of training samples can often lead to under-fitting and impede the model's capability to learn efficiently.

\textbf{What is the gain of combining the datasets?} To evaluate that, we constructed a  \textit{combined} dataset by combining the training, validation, and test sequences of the individual scenarios. This increases the size and diversity of the dataset. As shown in Fig.~\ref{fig:acc_bar_plot}, the model that is trained based on this combined dataset is generally achiving better than the models trained on the individual scenario datasets.

\begin{figure}[t]
	\centering
	\subfigure[Future-1]{\centering \includegraphics[width=0.49\linewidth]{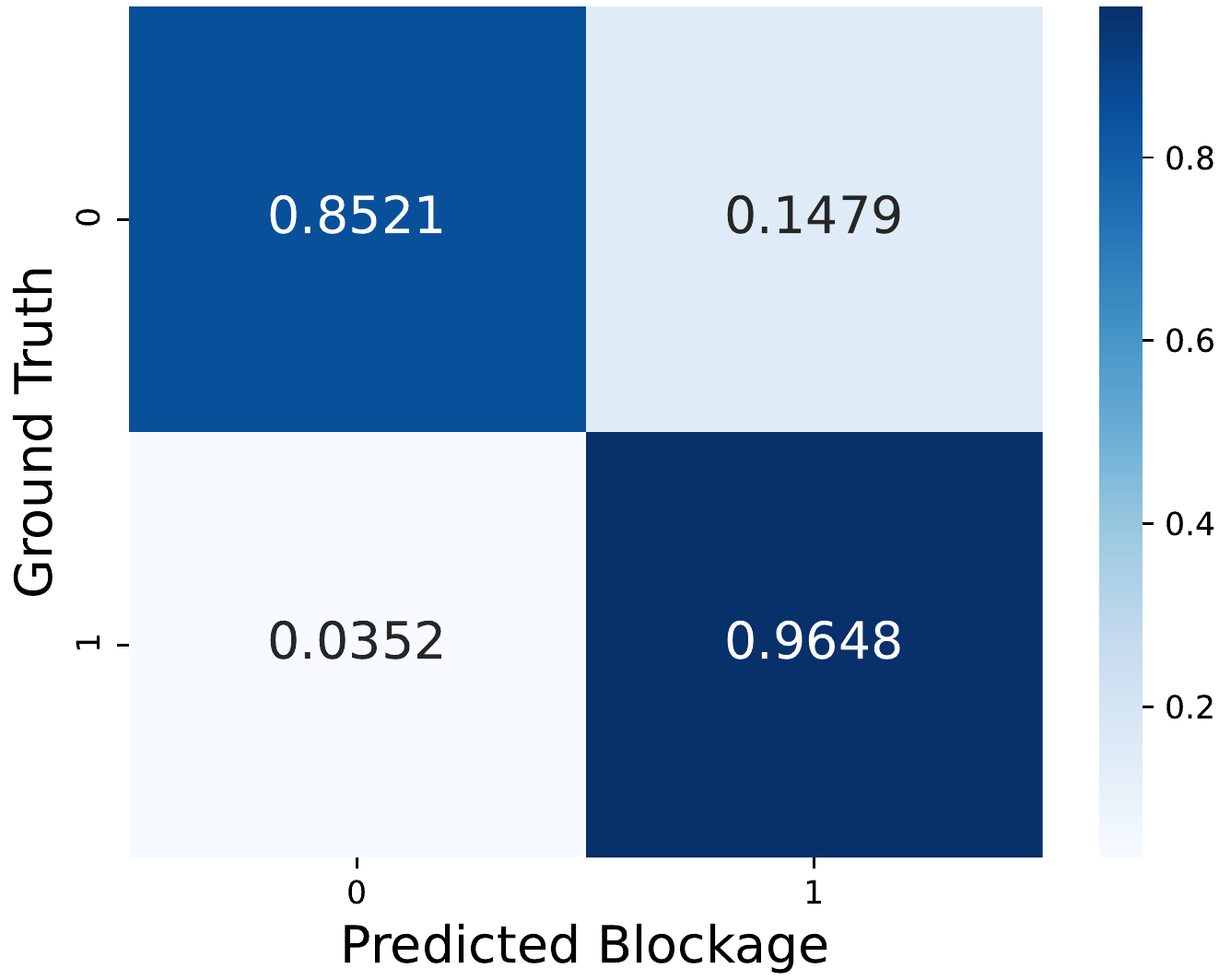}\label{fig:future_1_cf}}
	\subfigure[Future-10]{\centering \includegraphics[width=0.49\linewidth]{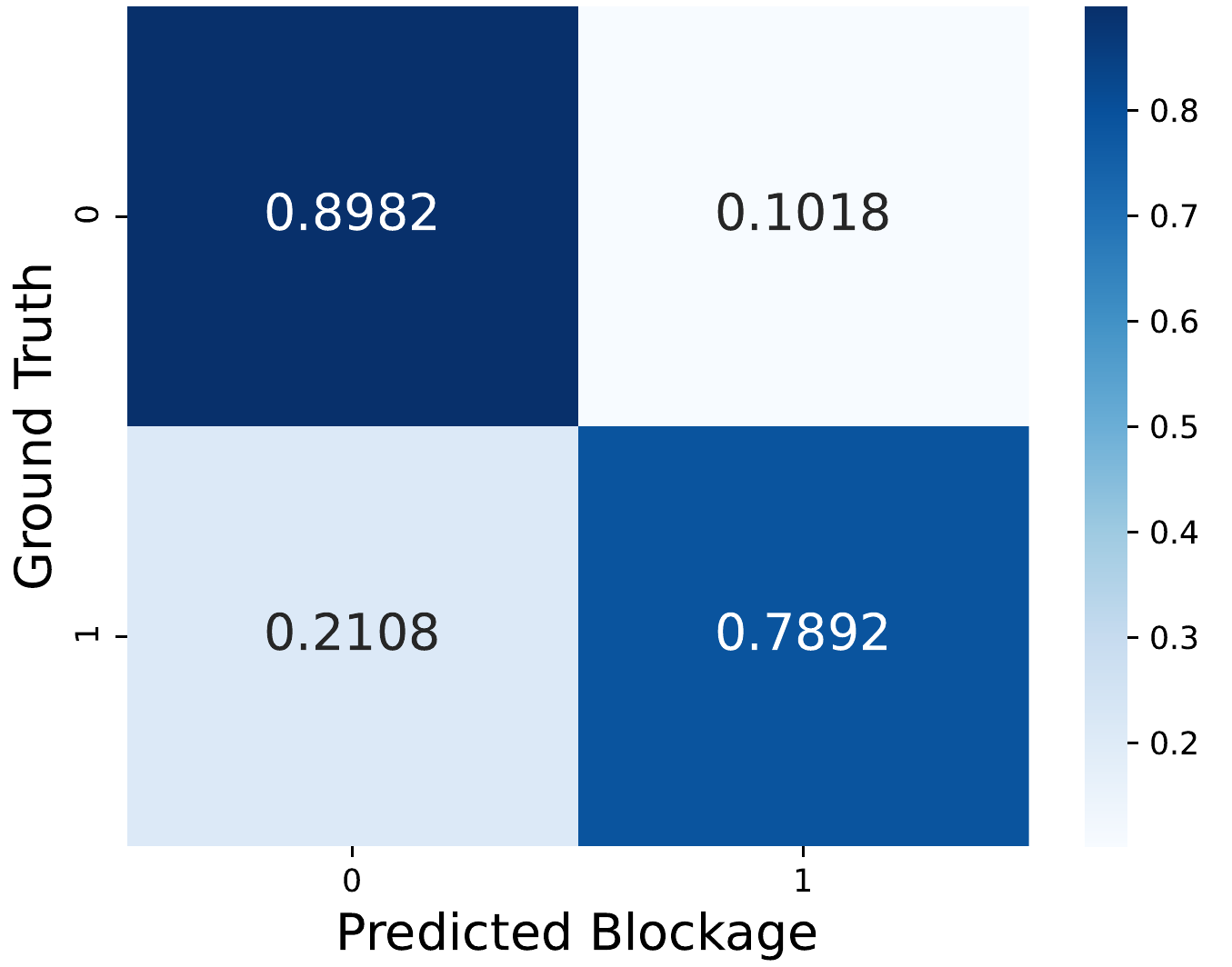}\label{fig:future_10_cf}}		

	\caption{This figure shows the confusion matrices for future-1 and future-10 blockage prediction interval (based on the combined dataset). It is observed that the model  efficiently differentiates between future LOS and NLOS links.  }
	\label{fig:combined_cf}
\end{figure}

\textbf{How early can a blockage be predicted?}  To answer this question, we evaluated the top1-accuracies and F1-scores for different future prediction window lengths (based on the combined dataset) in \figref{fig:combined_scene_score}. As shown in this figure, {the proposed approach achieves more than $90\%$ prediction accuracy till the future-5 prediction interval (an average of 600ms before the blockage happens)}. Even though the prediction accuracy and F1-score starts degrading after the future-5 instance, we observe that the model achieves almost $80\%$ accuracy for predicting up to the $10$th future instance. Accuracy being a holistic metric may not reflect the intricacies of the blockage prediction task. To develop a deeper insight into the model's performance, we plot the confusion matrices in Fig.~\ref{fig:future_1_cf} and Fig.~\ref{fig:future_10_cf}, for future-1 and future-10 combined predictions, respectively. The high precision of $96\%$ and $78\%$ for both cases further highlights the high efficiency of the proposed architecture in the future blockage prediction task.

%####################################################################################################
%####################################################################################################

\section{Conclusion}\label{sec:conc}

This paper explores the potential of leveraging visual sensory data for proactive blockage prediction in a mmWave communication system. We formulate the vision-aided blockage prediction problem and develop an efficient machine learning-based solution to predict future blockages. The key takeaways of evaluating our proposed vision-aided blockage prediction solution based on the large-scale real-world dataset, DeepSense, can be summarized as follows: (i) the vision-aided solution achieves high blockage prediction accuracy of more than $90\%$ for a shorter prediction window, i.e., for predicting future moving blockages that are within $600$ ms. (ii) For predicting further into the future (within one second), the proposed solution achieves an average prediction accuracy of more than $80\%$. These results highlight the potential gains of leveraging visual data in predicting future link blockages and enable proactive network management decisions. 

%\balance

\end{document}